\documentclass[submission, Proceedings]{SciPost}

\def\p{I\!\!P}

\usepackage{lineno}

\hypersetup{
    colorlinks,
    linkcolor={red!50!black},
    citecolor={blue!50!black},
    urlcolor={blue!80!black}
}

\usepackage[bitstream-charter]{mathdesign}
\DeclareSymbolFont{usualmathcal}{OMS}{cmsy}{m}{n}
\DeclareSymbolFontAlphabet{\mathcal}{usualmathcal}

\begin{document}

\begin{center}{\Large \textbf{
Prospects for diffractive dijet photoproduction at the EIC\\
}}\end{center}

\begin{center}
V.\ Guzey\textsuperscript{1} and
M.\ Klasen\textsuperscript{2$\star$}
\end{center}

\begin{center}
{\bf 1} National Research Center ``Kurchatov Institute'', Petersburg
 Nuclear Physics Institute (PNPI), Gatchina, 188300, Russia
\\
{\bf 2} Institut f\"ur Theoretische Physik, Westf\"alische Wilhelms-Universit\"at M\"unster, Wilhelm-Klemm-Stra\ss{}e 9, 48149 M\"unster, Germany
\\
* michael.klasen@uni-muenster.de
\end{center}

\begin{center}
\today
\end{center}


\definecolor{palegray}{gray}{0.95}
\begin{center}
\colorbox{palegray}{
  \begin{tabular}{rr}
  \begin{minipage}{0.1\textwidth}
    \includegraphics[width=22mm]{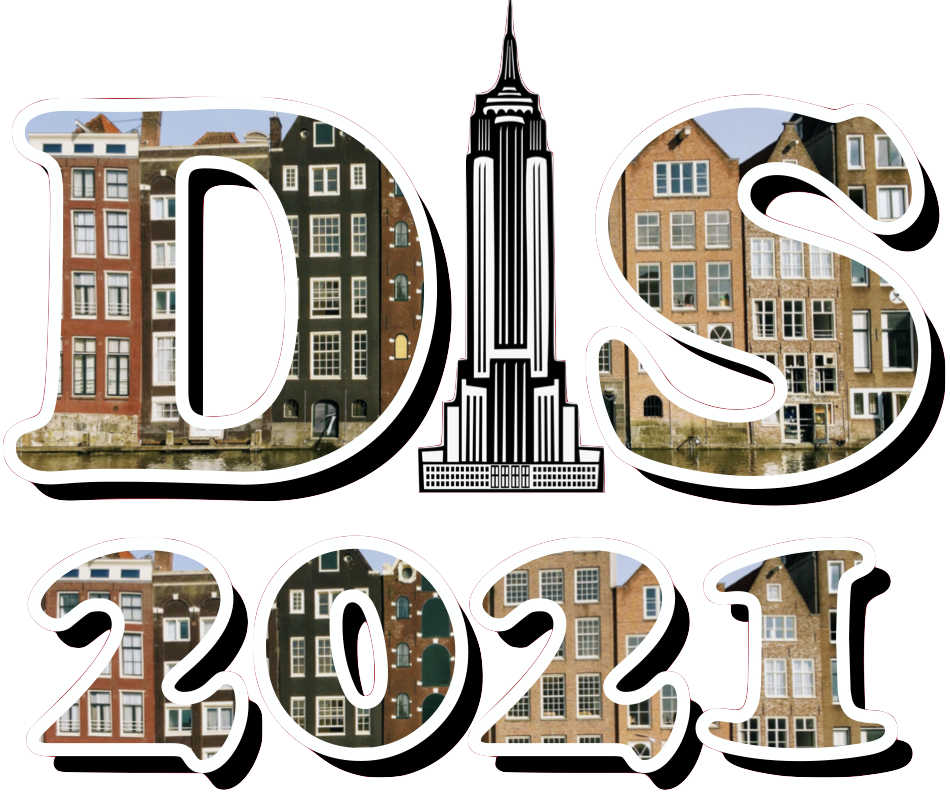}
  \end{minipage}
  &
  \begin{minipage}{0.75\textwidth}
    \begin{center}
    {\it Proceedings for the XXVIII International Workshop\\ on Deep-Inelastic Scattering and
Related Subjects,}\\
    {\it Stony Brook University, New York, USA, 12-16 April 2021} \\
    \doi{10.21468/SciPostPhysProc.?}\\
    \end{center}
  \end{minipage}
\end{tabular}
}
\end{center}

\section*{Abstract}
{\bf
We discuss the prospects of diffractive dijet photoproduction at the EIC to distinguish different fits of diffractive proton PDFs, different schemes of factorization breaking, to determine diffractive nuclear PDFs and pion PDFs from leading neutron production.
}

\section{Introduction}
\label{sec:1}
The HERA electron-proton collider, which operated at DESY Hamburg from 1992 to 2007, has provided us with a wealth of experimental information on the strong interaction and the quark and gluon structure of the proton \cite{Newman:2013ada}. In particular, parton density functions (PDFs) and the strong coupling constant $\alpha_s$ were determined with unprecedented precision, not only in inclusive deep-inelastic scattering (DIS), but also from jet production in DIS and photoproduction. It came as a surprise that a large fraction of events (e.g.\ 10-15\% in DIS) were diffractive, leaving the scattered proton intact or in an excited state. Using QCD \cite{Collins:1997sr} and assuming Regge factorization \cite{Ingelman:1984ns}, these processes could be described in terms of diffractive PDFs.

In photoproduction \cite{Klasen:2002xb}, factorization for diffractive processes turned out to be broken \cite{Klasen:2008ah}, similarly to observations at hadron colliders \cite{Klasen:2009bi}. The question whether this occurred only for photons resolved into their hadronic structure or also for direct photons (and thus globally) could not be answered, also because the two processes become related at next-to-leading order (NLO) of QCD and beyond \cite{Guzey:2016awf}.

The advent of a new electron-ion collider (EIC) at BNL raises hopes that a solution for this long-standing problem might be found, although we will see that this will not be easy due to the lower beam energies of the EIC compared to HERA and the ensuing dominance of direct photon processes \cite{AbdulKhalek:2021gbh}. However, the EIC will also provide access to so far completely unknown {\em nuclear} diffractive PDFs, which would then allow one to test theoretical models of nuclear shadowing and its relation to diffraction  \cite{Frankfurt:2011cs}. Finally, a leading neutron spectrometer would permit to resume studies of the virtual pion cloud in the nucleon and thus of the pion PDFs at low momentum fractions $x$ \cite{Klasen:2001sg}.

\section{Diffractive dijet photoproduction and HERA results}
\label{sec:2}

At HERA, electrons or positrons of 27.5 GeV energy collided with protons of first 820, then 920 GeV. Instrumentation with forward taggers allowed the study of both photoproduction and diffraction. Dijet events accounted for a large fraction of the cross section (e.g.\ 10-20\% in DIS) \cite{Aaron:2009vs}. The ratio of diffractive over inclusive dijets in photoproduction was 1-4\% \cite{Aaron:2010su}. 

\begin{figure}
\centering
\includegraphics[width=0.37\textwidth]{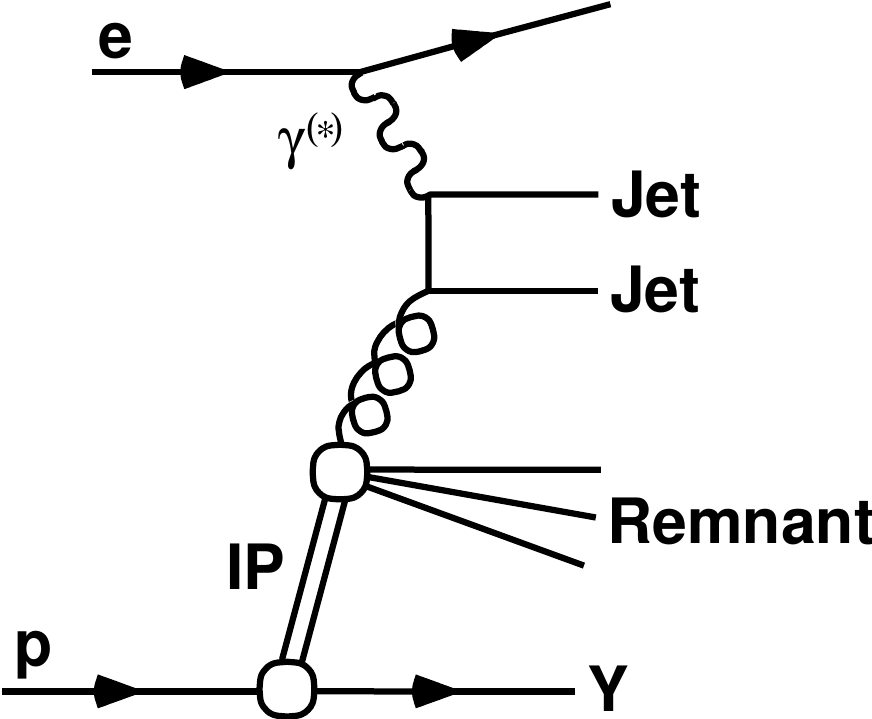}
\includegraphics[width=0.39\textwidth]{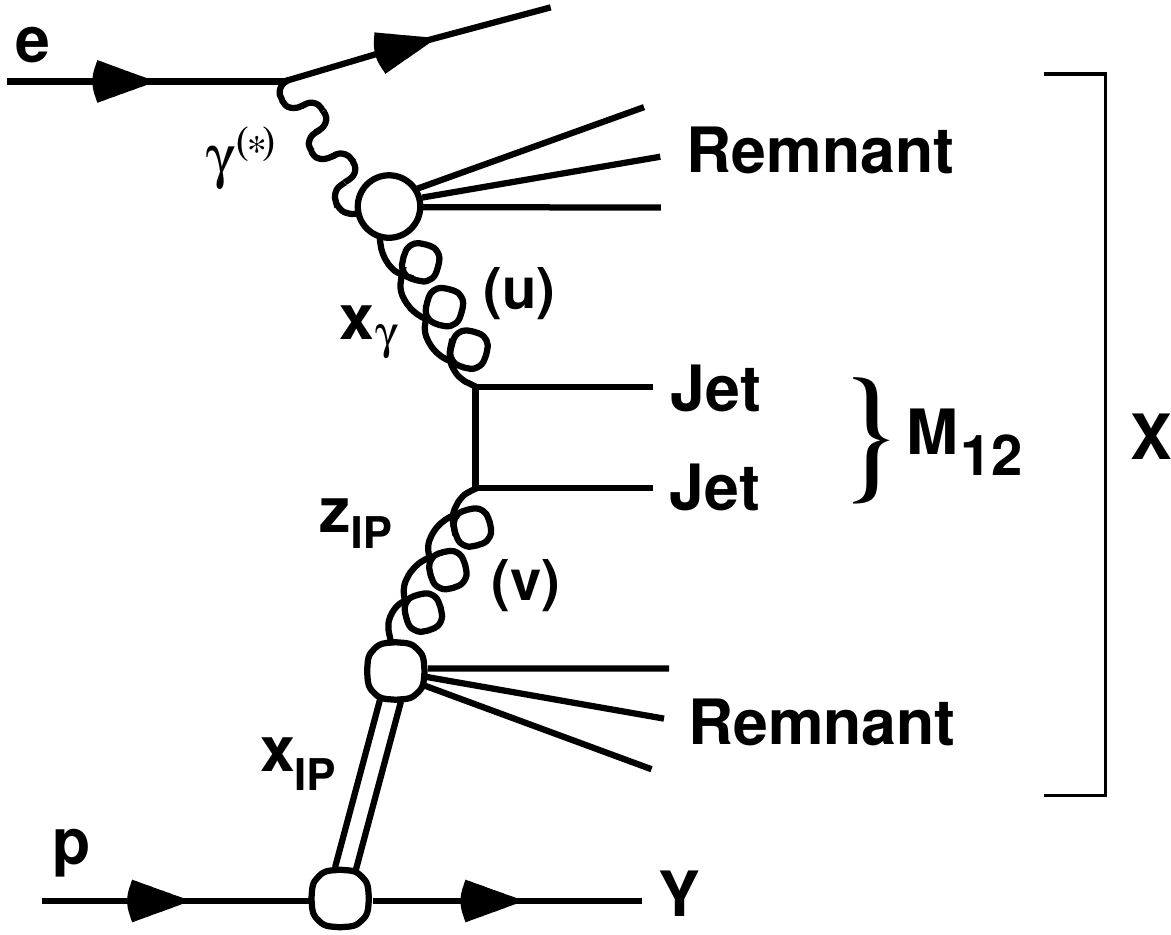}
\caption{Diffractive production of dijets with invariant mass
  $M_{12}$ in direct (left) and resolved (right) photon-pomeron collisions,
  leading to the production of one or two additional remnant jets.
  The hadronic systems $X$ and $Y$ are separated by the largest rapidity
  gap in the final state.}
\label{fig:1}
\end{figure}

The two mechanisms of direct and resolved diffractive dijet photoproduction are illustrated in Fig.\ \ref{fig:1}. In inclusive DIS (i.e.\ with virtual photons $\gamma^*$), diffractive PDFs were extracted by both the H1 \cite{Aktas:2006hy} and ZEUS \cite{Chekanov:2009aa} collaborations and tested successfully in DIS dijet and heavy quark production. When only a rapidity gap between the hadronic systems $X$ and $Y$ was imposed and the intact proton was not measured as in the H1 2006 Fits A and B, the normalization was larger by a factor of $1.23$ than when the diffractive PDFs were determined with a leading proton as in ZEUS 2009 Fit SJ. The H1 2006 Fits A and B differed mostly in their singular vs.\ regular gluon behavior at large observed parton momentum fraction $z_{\p}^{\rm obs.}$ in the exchanged pomeron $\p$.

\section{Electron-proton collisions at the EIC}
\label{sec:3}

At the EIC, electrons of energy 21...18 GeV will collide with protons of energy 100...275 GeV, leading to center-of-mass energies of $\sqrt{s}=92...141$ GeV. Since only longitudinal momentum fractions $y$ and $x_{\p}$, $x_{\gamma}^{\rm obs.}$ and $z_{\p}^{\rm obs.}$ are transmitted from the electron and proton to the photon and pomeron and their partonic constituents, the invariant mass $M_{12}$ of the hard dijet system is typically small and requires low kinematic cuts. From experience at HERA, the (asymmetric) cuts on the transverse momenta $p_{T1,(2)}$ can be as low as 5 (4.5) GeV \cite{Klasen:1995xe}. The $K$-factor of NLO/LO cross sections is about two for an anti-$k_T$ jet radius of $R=1$ except at the kinematic edges, where it can be larger \cite{Guzey:2020gkk}.

\begin{figure}
\centering
\includegraphics[width=0.4\textwidth]{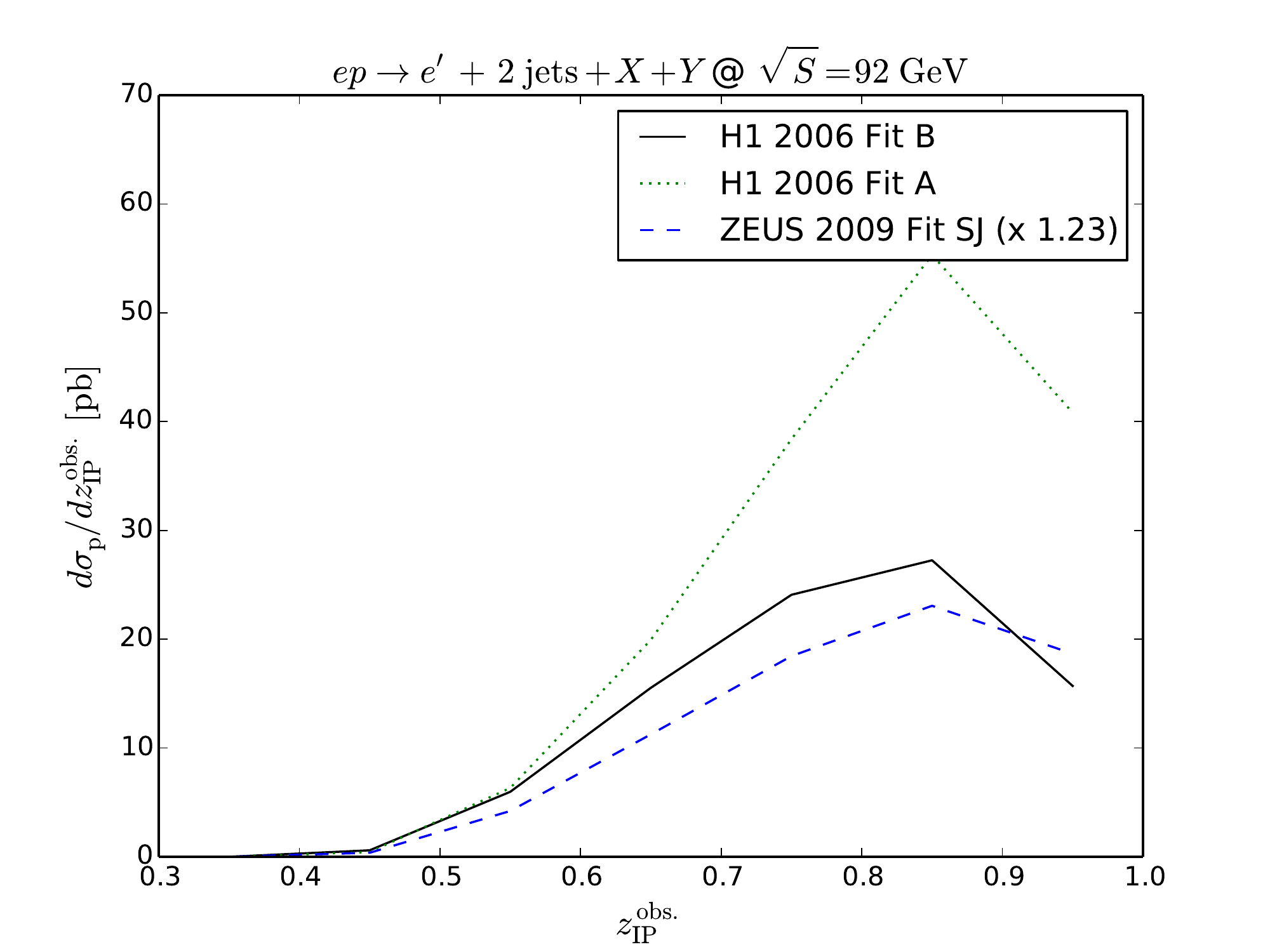}
\includegraphics[width=0.4\textwidth]{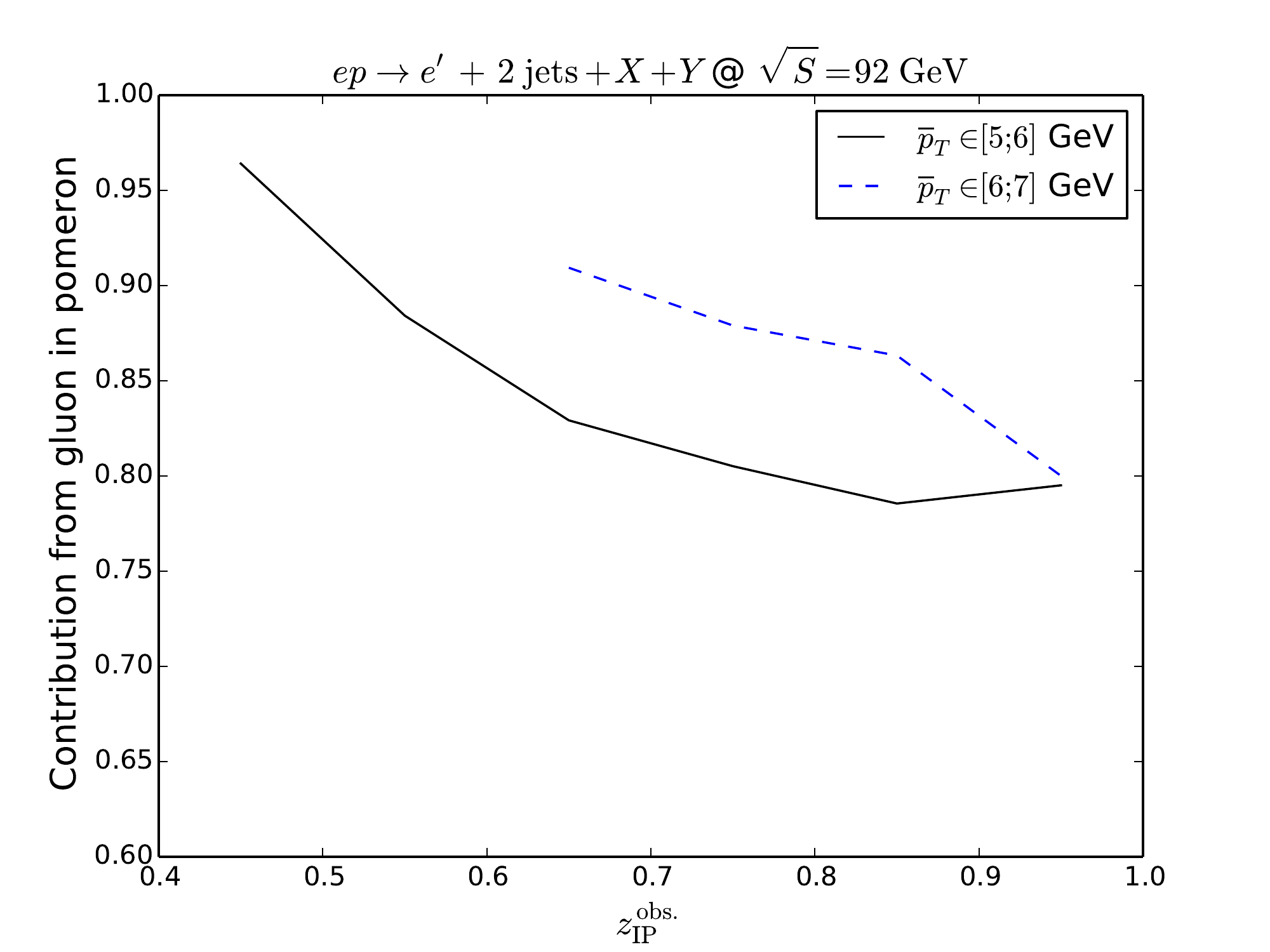}
\caption{Left: Distributions in the observed pomeron momentum fraction $z_{\p}^{\rm obs.}$ for three different
  sets of diffractive PDFs at NLO QCD. Right: Evolution in $Q^2\sim\bar{p}_T^2$ of the
  contribution from gluons in the pomeron. Here, $x_{\p}<0.03$ is imposed.}
\label{fig:2}
\end{figure}

As one can see in Fig.\ \ref{fig:2}, with 21 GeV $\times$ 100 GeV beams and the expected luminosity the two H1 2006 Fits A and B (which is similar to ZEUS 2009 Fit SJ) could be distinguished and the evolution of the diffractive PDFs tested, although only over relatively small ranges in $z_{\p}^{\rm obs.}$ and $Q^2\sim\bar{p}_T^2$.

\begin{figure}[h]
\centering
\includegraphics[width=0.4\textwidth]{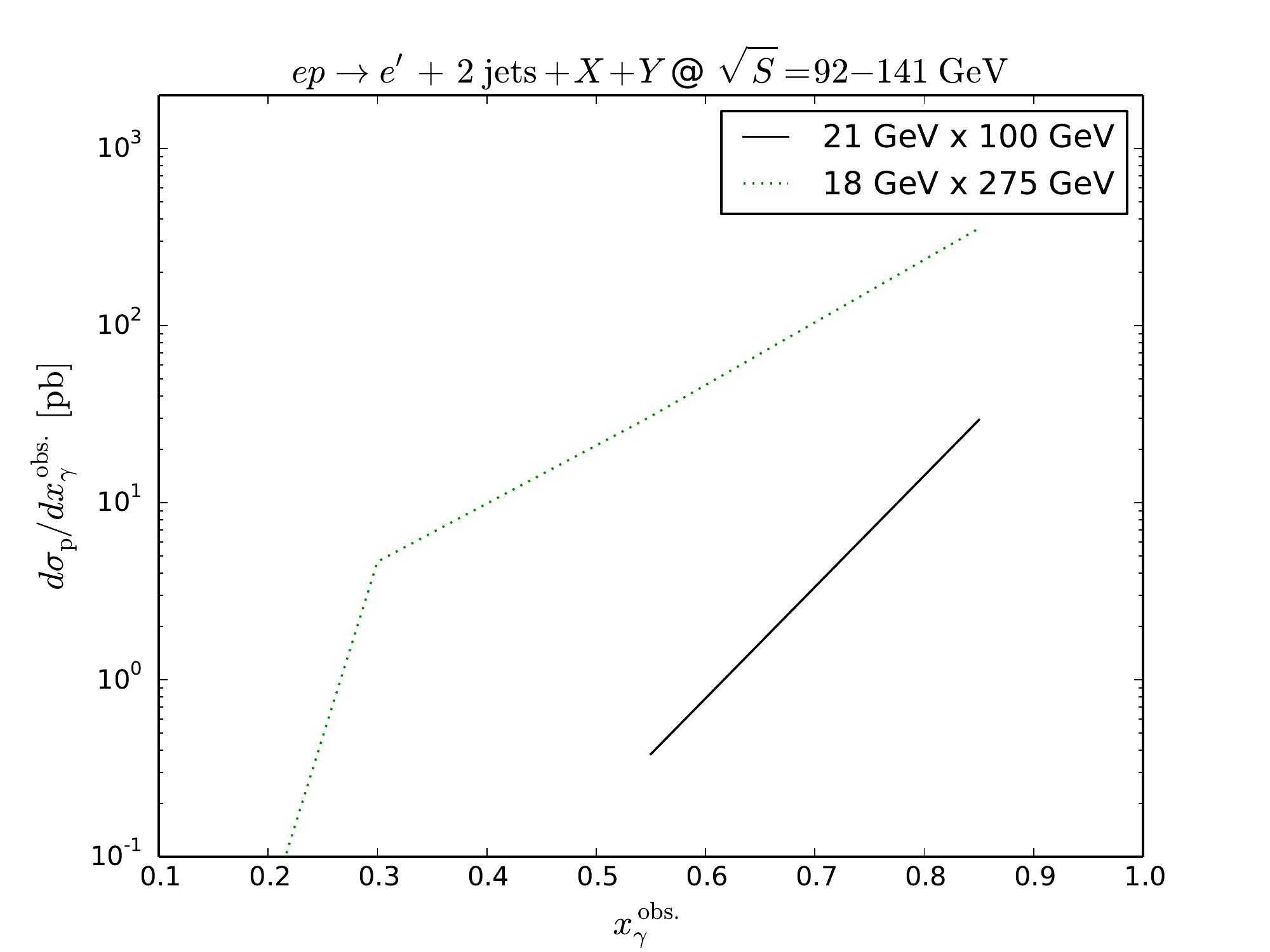}
\includegraphics[width=0.4\textwidth]{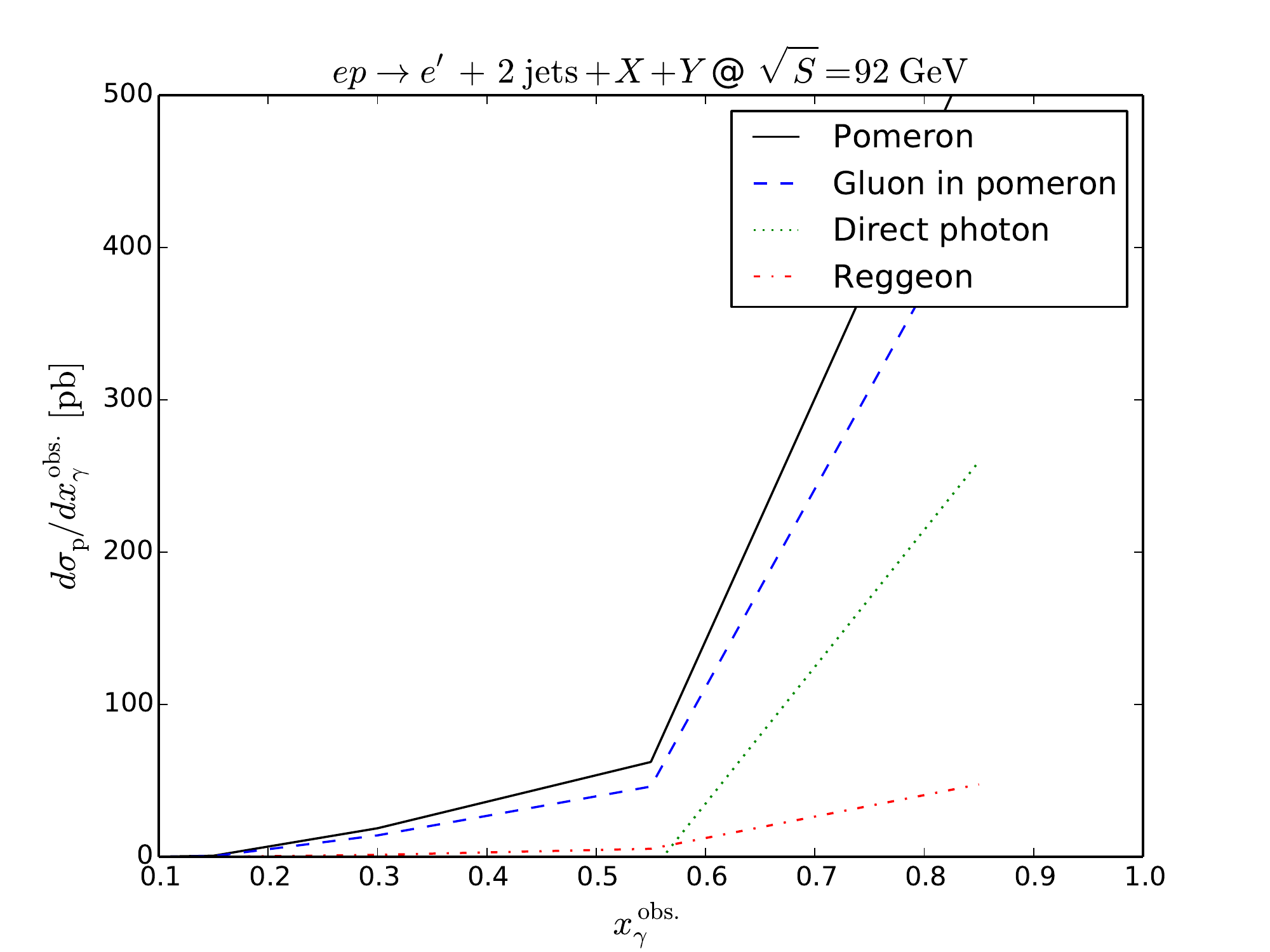}
\caption{Distributions in the observed photon momentum fraction $x_{\gamma}^{\rm obs.}$ in the 18 GeV $\times$ 275 GeV and 21 GeV $\times$ 100 GeV beam configurations (left) and with a higher cut on $x_{\p}<0.10$ (right).}
\label{fig:3}
\end{figure}

With 18 GeV $\times$ 275 GeV beams, the range in jet $p_T$ then extends to 12 rather than only 8 GeV and in $x_{\gamma}^{\rm obs.}$ and $z_{\p}^{\rm obs.}$ down to 0.2 rather than only 0.5, giving at least some access to the resolved photon contribution as shown in Fig.\ \ref{fig:3}. As also shown there, a similar effect can be obtained by increasing the cut on the pomeron momentum fraction $x_{\p}$ from 0.03 to 0.10, although this also increases the otherwise subleading reggeon contribution. These conditions give at least some hope to distinguish global from resolved-photon-only factorization breaking at the EIC \cite{Guzey:2020gkk}.

\section{Electron-ion collisions at the EIC}
\label{sec:4}

Coherent diffraction on nuclei can be selected with a large rapidity gap and by requiring that no neutrons are produced in the zero-degree calorimeter. At small $x$ nuclear diffractive PDFs are expected to be suppressed compared to the coherent sum of free nucleon diffractive PDFs due to nuclear shadowing. In the model of leading twist nuclear shadowing, it can be shown that \cite{Frankfurt:2011cs}
\begin{equation}
 f_{i/A}^{D}(z_{\p},Q^2,x_{\p})\approx A \cdot0.65\cdot f_{i/p}^{D}(z_{\p},Q^2,x_{\p}).
\end{equation}
In addition, since the nuclei stay intact, the diffractive free nucleon PDFs obtained from processes including diffractive dissociation like H1 2006 Fits A and B must be divided by a factor of 1.23.
\begin{figure}[h]
\centering
\includegraphics[width=0.4\textwidth]{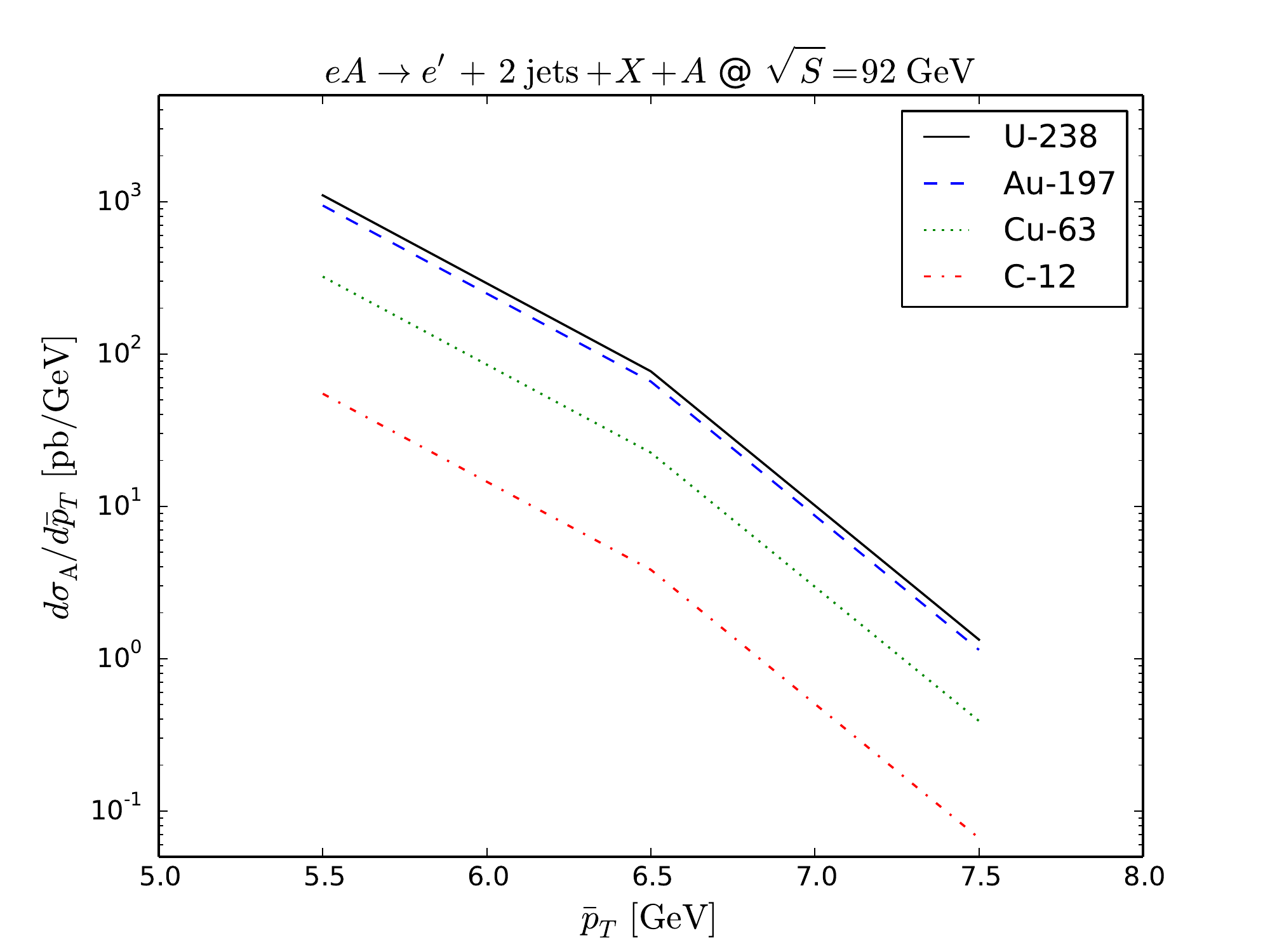}
\includegraphics[width=0.4\textwidth]{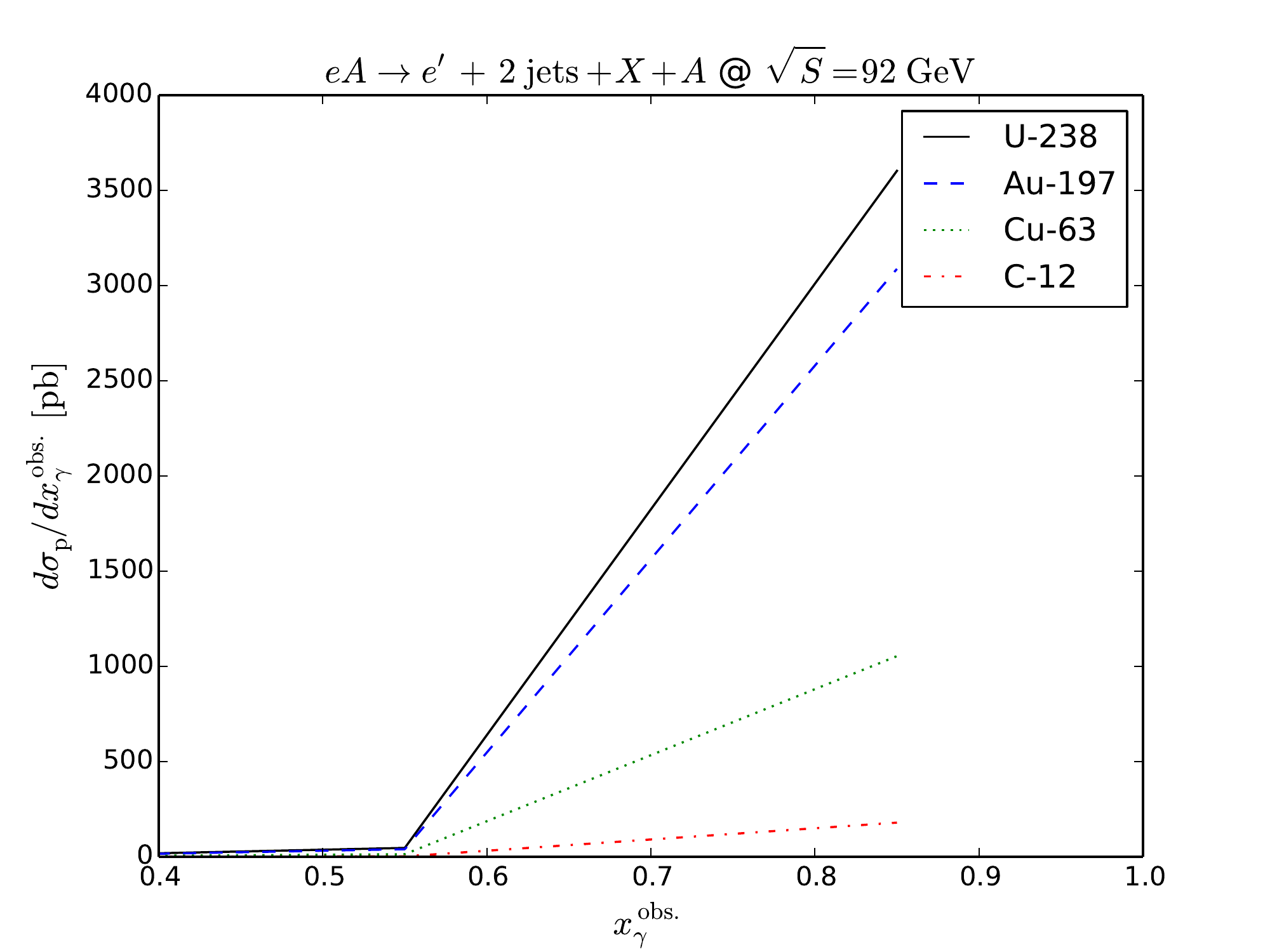}
\includegraphics[width=0.4\textwidth]{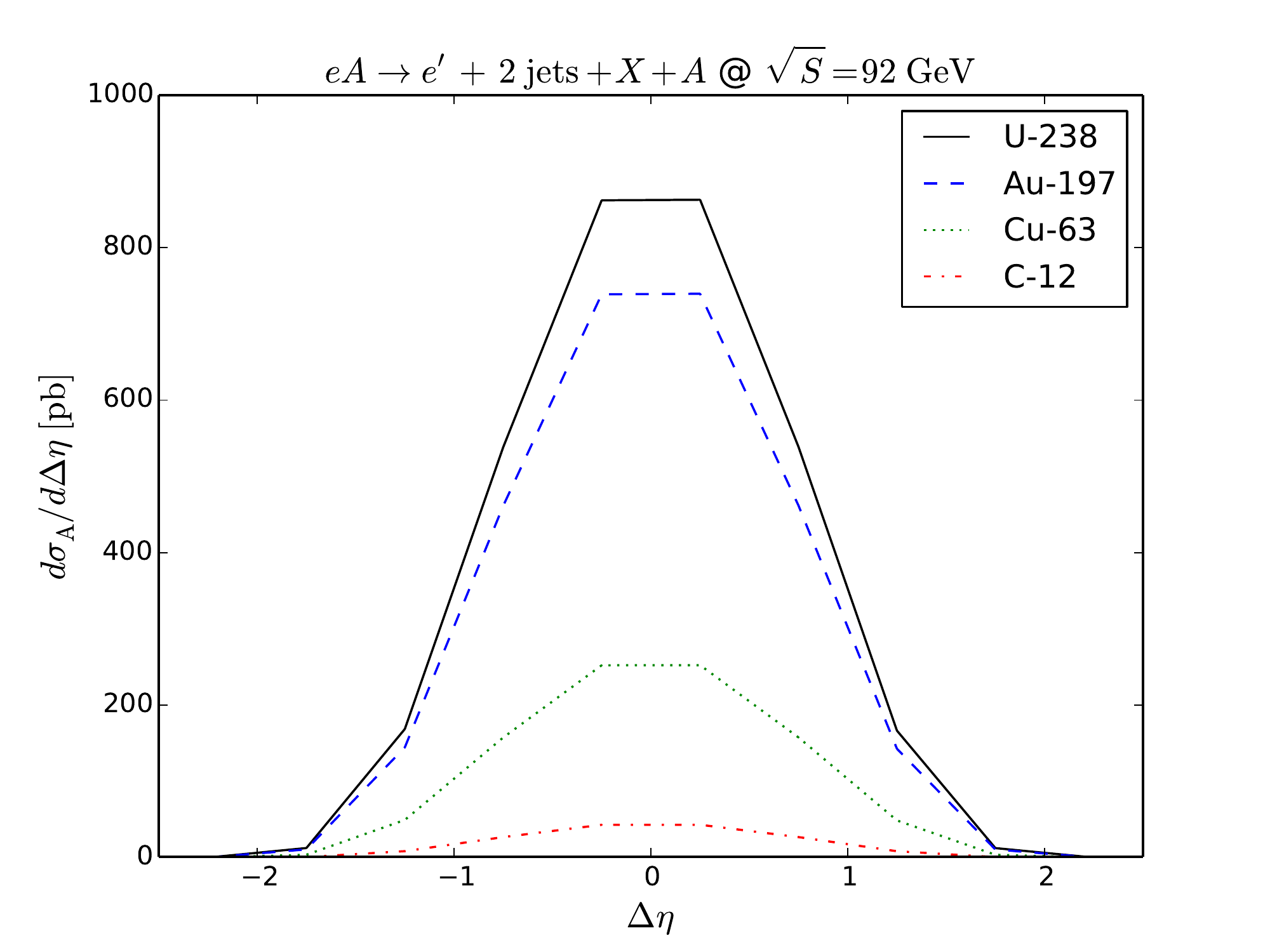}
\includegraphics[width=0.4\textwidth]{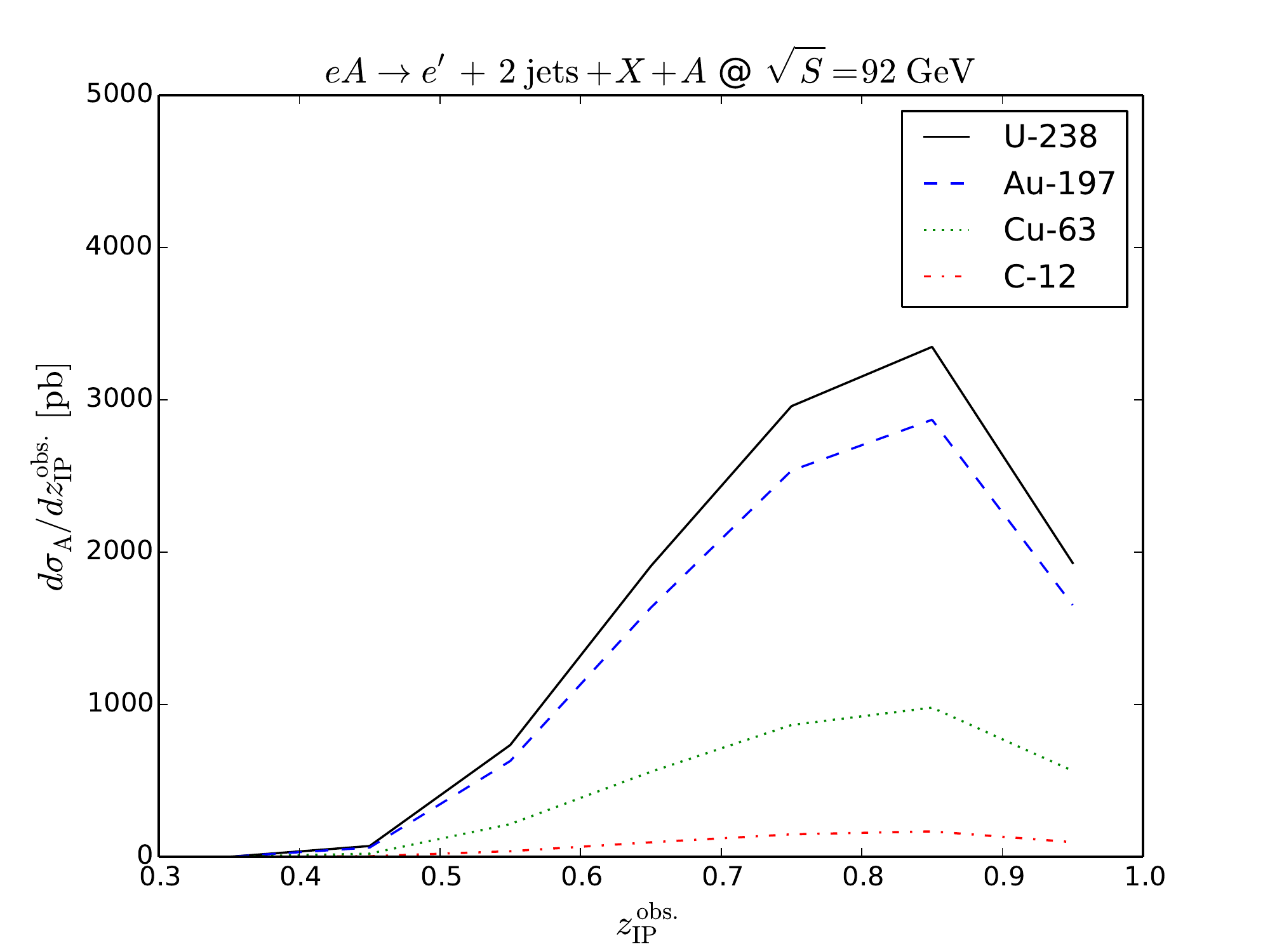}
\caption{NLO QCD cross sections for coherent  diffractive dijet photoproduction 
 $eA \to e^{\prime}+2\ {\rm jets}+ X+A$
 with various nuclear beams 
 and a center-of-mass energy per nucleon
 of $\sqrt{s}=92$ GeV at the EIC. The cross sections are shown as functions of 
 the jet average transverse momentum (top left) and rapidity difference (bottom left) 
 as well as the longitudinal
   momentum fractions in the photon (top right)
   and pomeron (bottom right).}
\label{fig:4}
\end{figure}
Our NLO predictions for diffractive dijet photoproduction on different nuclei are shown in Fig.\ \ref{fig:4}. Scaling these predictions with a global suppression factor of 0.5 or with a factor 0.04 for resolved-only factorization breaking as expected from the commonly used two-state eikonal model\cite{Khoze:2000wk} leads to very similar results, making a distinction of these two schemes again difficult \cite{Guzey:2020gkk}.

\section{Conclusion}
\label{sec:5}

Identifying the mechanism of factorization breaking in diffractive photoproduction is a desideratum of the HERA program which should be addressed at the EIC. However, we have seen that this will not be easy due to the lower EIC center-of-mass energy and the ensuing dominance of direct photon processes. Access to nuclear diffractive PDFs is, however, a novel unique opportunity at the EIC, where in particular the connection of diffraction and nuclear shadowing can be addressed experimentally.

\section*{Acknowledgements}

Financial support by the DFG through the grant KL 1266/9-1 within
the framework of the joint German-Russian project ``New constraints on nuclear
parton distribution functions at small $x$ from dijet production in $\gamma A$
collisions at the LHC" is gratefully acknowledged. VG's research is also
supported in part by RFBR, research project 17-52-12070.

\bibliography{klasen.bib}

\nolinenumbers

\end{document}